\documentclass[twocolumn,pra,aps,showpacs]{revtex4}
\usepackage{graphicx}
\usepackage{amsmath}
\usepackage{amssymb}

\begin{document}

\title{Efficient measurements, purification, and bounds on the mutual information}


\author{Kurt Jacobs}

\affiliation{Centre for Quantum Computer Technology, Centre for 
Quantum Dynamics, School of Science, Griffith University, Nathan 4111, 
Brisbane, Australia}

\begin{abstract}
When a measurement is made on a quantum system in which classical information is encoded, the measurement reduces the observers average Shannon entropy for the encoding ensemble. This reduction, being the {\em mutual information}, is always non-negative. For efficient measurements the state is also purified; that is, on average, the observers von Neumann entropy for the state of the system is also reduced by a non-negative amount. Here we point out that by re-writing a bound derived by Hall [Phys.\ Rev.\ A {\bf 55}, 100 (1997)], which is dual to the Holevo bound, one finds that for efficient measurements, the mutual information is bounded by the reduction in the von Neumann entropy. We also show that this result, which provides a physical interpretation for Hall's bound, may be derived directly from the Schumacher-Westmoreland-Wootters theorem [Phys.\ Rev.\ Lett.\ {\bf 76}, 3452 (1996)]. We discuss these bounds, and their relationship to another bound, valid for efficient measurements on pure state ensembles, which involves the subentropy.
\end{abstract}


\maketitle

In what follows we will be concerned with the situation in which one 
observer, the sender, transmits information to another observer, the 
receiver, by encoding that information in a quantum system and having 
the receiver make a subsequent measurement on the system. It is useful 
at this point to define all our terminology and notation regarding this 
information transmission process. 

To encode the information in the quantum system the sender uses an 
alphabet consisting of a set of possible states, and 
prepares the system in one of these states, $\rho_{i}$, with 
probability $P_{i}$.  The set of states, along with their respective 
probabilities is referred to as the {\em encoding}, or the 
{\em ensemble}, and we will denote 
it by $\varepsilon \equiv \{P_{i},\rho_{i}\}$.  When the system has 
been prepared by the sender, the state-of-knowledge of the receiver 
regarding the system is $\rho = \sum_{i}P_{i}\rho_{i}$.  We will 
always denote the dimension of the system used for encoding by $N$, 
and we will refer to $\rho$ as the {\em ensemble state}.

The measurement made by the receiver is described by a set of 
operators, $A_{j}$, such that $\sum_{j}A_{j}^{\dagger}A_{j} = 
I$~\cite{Schumacher,Krauss,introMeas}.  We will denote the measurement 
by ${\cal M} \equiv \{A_{j}\}$, and where convenient denote the 
operators $A_{j}^{\dagger}A_{j}$ as $E_{j}$.  For efficient measurements, 
with which we will be concerned in the following unless otherwise stated,
each of the operators 
$A_{j}$ corresponds to a measurement outcome, and the outcomes are 
therefore labeled by $j$.  The final state of the system from the 
point of view of the observer, having obtained the outcome $j$, is 
$\rho'_{j} = A_{j}\rho A_{j}^{\dagger}/Q_{j}$, where $Q_{j} = 
\mbox{Tr}[E_{j}\rho]$ is the probability that outcome 
$j$ will result.  For clarity we will denote probability densities 
over $i$ as $P$, and those over $j$ as $Q$.

The amount of information transmitted to the receiver in the process 
of preparation and measurement, which we will refer to as $\Delta 
I_{\mbox{\scriptsize i}}$, is given by the mutual information, 
$H(I\!\!  : \!\!  J)$, between the preparation, indexed by $i$, and 
the outcomes, indexed by $j$~\cite{Shannon}.  Thus
\begin{equation}
  \Delta I_{\mbox{\scriptsize i}} =	H(I\!\! : \!\! J) = 
                                    H[P_i] - \sum_jQ_j H[P(i|j)] ,
\end{equation}
where $H$ is the Shannon entropy, and $P(i|j)$ is the receiver's 
probability density for the prepared state, after having received 
outcome $j$.  (That is, the receiver's final state-of-knowledge about 
which state was initially prepared.)  The mutual information is 
thus the average difference between the receiver's initial information 
about the preparation, and her final information after the 
measurement.  We denote this by $\Delta I_{\mbox{\scriptsize i}}$ to 
reflect this fact, with the subscript indicating that it constitutes 
information about the {\em initial} preparation.  The maximum of 
$\Delta I_{\mbox{\scriptsize i}}$ over all measurements, for a fixed 
encoding, is referred to as the {\em accessible information} of the 
encoding~\cite{accI}, and we will denote this by $\Delta 
I_{\mbox{\scriptsize acc}}$.

The celebrated Holevo bound provides a limit to the accessible 
information of an encoding~\cite{Holevo73,YO93,FC94}.  The Holevo 
bound is
\begin{equation} 
   \Delta I_{\mbox{\scriptsize i}} \leq S[\rho] - 
   \sum_{i}P_{i}S[\rho_{i}] \equiv \chi(\varepsilon) ,
\end{equation}
where $S[\rho]$ denotes the von Neumann entropy of $\rho$.  
 
One can also a problem which may be viewed as being complementary to that of finding the accessible information; that of obtaining the maximum of $\Delta I_{\mbox{\scriptsize i}}$ given that it is the receiver's measurement and the ensemble state $\rho$ which are fixed, and it is instead the sender which has the ability to use any encoding consistent with $\rho$. Hall has shown that it is possible to use Holevo's bound, along with a duality relation between encodings and measurements, which he refers to as {\em source duality}, to derive a bound on $\Delta I_{\mbox{\scriptsize i}}$ for this case.  Hall's dual Holevo bound is~\cite{Hall}
\begin{equation}
   \Delta I_{\mbox{\scriptsize i}} \leq S[\rho] - \sum_{j}Q_{j}
                           S\left[ \frac{\sqrt{\rho}E_{j}\sqrt{\rho}}{Q_{j}}\right] .
\end{equation}
The Holevo bound and (as we will show) the dual Holevo bound, may both be derived 
directly from the more general bound obtained by Schumacher, Westmoreland and Wootters 
in 1996~\cite{SWW96}. We state this theorem now, and will return to it later.

\vspace{1mm} {\em Theorem} [Schumacher-Westmoreland-Wootters]: The 
information transmitted from sender to receiver, $\Delta 
I_{\mbox{\scriptsize i}}$, when the sender uses the encoding 
$\varepsilon$, and the receiver uses measurement ${\cal M}$, is 
bounded such that
\begin{eqnarray} 
   \Delta I_{\mbox{\scriptsize i}} & \leq & S[\rho] - \sum_{i}P_{i}S[\rho_{i}] 
\nonumber \\ 
                                   & & - \sum_{j}Q_{j}\left[ S[\rho'_{j}] - 
                                         \sum_{i} P(i|j) S[\rho'_{ji}] 
                                         \right]
   \label{sww} 
\end{eqnarray}
where all quantities are as defined above, and a new quantity is 
introduced, being $\rho'_{ji}$, which is the final state that the 
receiver {\em would} have had, {\em if} she knew that the initial state 
was $\rho_{i}$.  Thus $\rho'_{ji} = A_{j}\rho_{i} 
A_{j}^{\dagger}/Q(j|i)$, where $Q(j|i)$ is naturally the probability 
density for the measurement outcomes, given that the initial state is 
$\rho_{i}$. Because of the final term on the right hand side of this inequality, to which we will return later, this bound is, in general, stronger than the Holevo bound.

While $\Delta I_{\mbox{\scriptsize i}}$ quantifies the information which the observer obtains about the initial preparation, there exists another quantity which can be said to characterize the average amount of information which the receiver obtains about the final state which she is left with after the measurement~\cite{DJJ,FJ}. We will denote this by $\Delta I_{\mbox{\scriptsize f}}$, the expression for which is 
\begin{equation} 
   \Delta I_{\mbox{\scriptsize f}} = S[\rho] - \sum_{j}Q_{j}S[\rho'_{j}] .
\end{equation}
This is the average difference between the receiver's initial von Neumann entropy of the quantum system, and her final von Neumann entropy. This quantity is useful when considering quantum state preparation and, more generally, quantum feedback control~\cite{DJJ}.  

While we have introduced $\Delta I_{\mbox{\scriptsize i}}$ and $\Delta I_{\mbox{\scriptsize f}}$ in terms of initial states and final states, the former is not really any more connected with initial states than it is with final states, since the Shannon entropy of the ensemble after measurement is independent of whether it is written in terms of the initial states or the final ones. A more fundamental difference between $\Delta I_{\mbox{\scriptsize i}}$ and $\Delta I_{\mbox{\scriptsize f}}$ is that the former is the average change in the observers Shannon entropy regarding the ensemble, where as the latter is the average change in the observers von Neumann entropy regarding the overall state of the quantum system. That is, $\Delta I_{\mbox{\scriptsize i}} = \langle \Delta H(\varepsilon) \rangle$ and $\Delta I_{\mbox{\scriptsize f}} = \langle \Delta S(\rho(\varepsilon)) \rangle$. 

We now ask, is there a relationship between the two kinds of information gathering, $\Delta I_{\mbox{\scriptsize i}}$ and $\Delta I_{\mbox{\scriptsize f}}$? It turns out that the answer to this question is yes: the former is bounded by the latter (that is, $\langle\Delta H\rangle \leq \langle\Delta S\rangle$), and this is readily shown by observing that it is actualy an alternative form for Hall's dual to Holevo's bound. To do this one notes that if we use the polar decomposition theorem~\cite{Schatten} to write $A_{j}=U_{j}\sqrt{E_{j}}$, where $U_{j}$ is unitary, and define $B_{j} = \sqrt{E_{j}\rho}$, then the Hermitian operators which appear in the dual bound are
\begin{equation}
   \sqrt{\rho}E_{j}\sqrt{\rho} = B_{j}^{\dagger}B_{j} ,
\end{equation}
while the final states are
\begin{equation}
   Q_{j} \rho_{j}' = U_{j} B_{j} B_{j}^{\dagger} U_{j}^{\dagger}.
\end{equation}
But $B_{j}^{\dagger}B_{j}$ and $B_{j} B_{j}^{\dagger}$ have the same eigenvalues~\cite{MO}. Thus since the von Neumann entropy is only a function of the eigenvalues, we can replace $\sqrt{\rho}E_{j}\sqrt{\rho}$ with $Q_{j} \rho_{j}'$ in the original expression for the dual bound, and the result is
\begin{equation}
  \Delta I_{\mbox{\scriptsize i}} \leq \Delta I_{\mbox{\scriptsize f}} .
  \label{mb}
\end{equation}
One can interpret this as saying that an observer cannot learn more about the classical information encoded in a quantum system than she learns about the state of the quantum system. This provides a physical interpretation for Hall's dual bound. It is also worth noting, as was pointed out by Hall~\cite{Hall}, that this bound is only saturated when all the operators $E_{j}$ commute.
 
The above result may also be obtained from the SWW theorem. To do this one first re-writes the second and fourth terms of the RHS of Eq.(\ref{sww}), using the fact that $Q_{j}P(i|j) = P_{i}Q(j|i)$:
\begin{eqnarray}
  & & - \sum_{i}P_{i}S[\rho_{i}] + \sum_{j}Q_{j}\sum_{i} P(i|j) 
  S[\rho'_{ji}] \nonumber \\ 
                & = & - \sum_{i}P_{i} 
                      \left[ S[\rho_{i}] - Q(j|i)S[\rho'_{ij}] \right]
                      \nonumber \\
                & = & - \sum_{i}P_{i} \Delta I_{\mbox{\scriptsize f}i} ,                                        
\end{eqnarray}
where $\Delta I_{\mbox{\scriptsize f}i}$ is the information that would 
have been obtained about the final state {\em if} the initial state had 
been $\rho_{i}$. This gives
\begin{eqnarray} 
   \Delta I_{\mbox{\scriptsize i}} & \leq & S[\rho] 
                               - \sum_{i}P_{i} \Delta I_{\mbox{\scriptsize f}i}
                               - \sum_{j}Q_{j}S[\rho'_{j}] .
   \label{eqx}
\end{eqnarray}
Now, since Nielsen has shown that $\Delta I_{\mbox{\scriptsize f}}$ is 
always non-negative~\cite{Nielsen} (see also \cite{FJ}), the RHS is 
maximized when the $\Delta I_{\mbox{\scriptsize f}i}$ are zero for all 
$i$.  Since this is true for all pure state ensembles, the result is 
the bound given in Eq.(\ref{mb}). 

One consequence of Eq.(\ref{mb}) is that, if we choose an ensemble which 
has the maximal accessible information for a fixed $\rho$, we can only 
obtain all this information if all the final states are pure.  As SWW 
point out in their paper, measurements which leave the final state 
impure, leave some information in the system.  That is, if the final 
state is mixed, in general it depends on the initial ensemble, and as 
a result subsequent measurements can obtain further information about 
the initial preparation, whereas this is not possible if the final 
state is pure.

For a given $\rho$ not all ensembles have an accessible information 
equal to $S[\rho]$.  We may ask then, if it is possible for 
measurements which leave the final state impure to extract all the 
accessible information from {\em these} encodings. In fact, this is only 
possible if the encoding satisfies special conditions; in 
general, incomplete measurements will not even extract the accessible 
information from an ensemble. To see this, consider the final states, 
$\rho'_{j}$, which result from the measurement.  Each of these consists 
of an ensemble, $\varepsilon_{j}$ over the states $\rho_{i|j}$, introduced 
above.  In particular,
\begin{equation}
   \rho'_{j} = \sum_{i} P(i|j) \rho_{i|j} .
\end{equation}
Since these ensembles consist of states indexed by $i$, they can, in 
general, be measured to obtain further information about the initial 
preparation. Since the accessible information is the maximal amount 
of information that can be obtained about $i$ by making measurements, 
we have the inequality
\begin{equation}
   \Delta I_{\mbox{\scriptsize i}}(\varepsilon,{\mathcal M}) \leq 
      \Delta I_{\mbox{\scriptsize acc}}(\varepsilon) 
   - \sum_{j} Q_{j} \Delta I_{\mbox{\scriptsize acc}}(\varepsilon_{j}).
   \label{accb}
\end{equation}
Thus, $\Delta I_{\mbox{\scriptsize i}}(\varepsilon,{\mathcal M})$ can 
only be equal to $\Delta I_{\mbox{\scriptsize acc}}(\varepsilon)$ if 
the $\Delta I_{\mbox{\scriptsize acc}}(\varepsilon_{j})$ are zero for 
all $j$.  If $\rho'_{j}$ is pure, then $\Delta I_{\mbox{\scriptsize 
acc}}(\varepsilon_{j})$ is zero.  If $\rho'_{j}$ is not pure, then the 
accessible information of $\varepsilon_{j}$ is only zero if, for any 
given $j$, the $\rho_{i|j}$ are the {\em same} for all $i$.  A little 
algebra shows that this is only true if
\begin{equation}
   P_{A_{j}}\rho_{i}P_{A_{j}} - \alpha_{ikj} P_{A_{j}}\rho_{k}P_{A_{j}} = 0, 
            \;\;\; \forall \; i,\! k \; ,
   \label{eqspec}
\end{equation}
for some non-negative real numbers $\alpha_{ikj}$, where $P_{A_{j}}$ 
is a projector onto the support of the operator $A_{j}$.  This means 
that for a measurement to extract all the accessible information, all 
the coding states $\rho_{j}$ must be identical, up to a multiplier, on 
the supports of the operators $A_{j}$, separately for every $j$. 

For pure-state ensembles it is easy to see the effect of the 
conditions given by Eq.(\ref{eqspec}).  Consider merely the $j$ for 
which the corresponding $A_{j}$ has the support with the largest 
dimension, and call this dimension $M_{\mbox{\scriptsize max}}$.  Then 
the effect of Eq.(\ref{eqspec}) for this $j$ alone is simply to limit 
the dimension of the space from which the pure states in the ensemble 
can be drawn to $N-M_{\mbox{\scriptsize max}}+1$.  The accessible 
information of pure-state ensembles which satisfy Eq.(\ref{eqspec}) is 
therefore bounded by $\ln(N-M_{\mbox{\scriptsize max}}+1)$.

As was noted by SWW, the expression in the square brackets in 
Eq.(\ref{sww}) is the Holevo $\chi$ quantity for the ensemble 
$\varepsilon_{j}$ which results from measurement outcome $j$. Thus 
their bound may be written as
\begin{equation}
   \Delta I_{\mbox{\scriptsize i}} \leq \chi[\varepsilon] 
                                   - \sum_{j}Q_{j}\chi[\varepsilon_{j}] .
   \label{sww2}
\end{equation}
Now, $\chi[\varepsilon_{j}]$ is the Holevo bound on the information 
that the receiver could extract when making a subsequent measurement 
after obtaining result $j$.  The SWW bound is therefore very 
interesting because it shows that, if the initial ensemble 
$\varepsilon$ is chosen so that its accessible information is maximal 
(ie.  equal to $\chi(\varepsilon)$), then the information obtained by 
an incomplete measurement will be reduced by the {\em maximal} amount 
of information which could be accessible from the final ensembles 
$\varepsilon_{j}$, and not merely the {\em actual}  
information available in these ensembles, which would imply the bound 
given in Eq.(\ref{accb}).  In 
general, therefore, there is a gap between the information lacking in 
an incomplete measurement, and that which can be recovered by 
subsequent measurements.  It is natural to ask therefore if this is 
true for all ensembles.  That is, whether the inequality in 
Eq.(\ref{accb}) can be strengthened by replacing the final term on the 
RHS by the final term in the RHS of Eq.(\ref{sww2}).  However, this is 
not the case.  As the following theorem shows, for pure-state 
ensembles a tight upper bound is obtained by replacing the final term 
in Eq.(\ref{accb}) by the average of the subentropies of the final 
states~\cite{JRW}, rather than the corresponding Holevo quantities.

\vspace{1mm} {\em Theorem}: For an initial pure-state ensemble 
$\varepsilon$, and a general measurement ${\mathcal M}$, one has
\begin{equation}
   \Delta I_{\mbox{\scriptsize i}} \leq  \Delta I_{\mbox{\scriptsize 
   acc}} [\varepsilon] - \sum_{j}Q_{j}Q[\rho'_{j}],
   \label{bsub} 
\end{equation}
where $Q[\cdot]$ is the subentropy as defined by Jozsa, Robb and 
Wootters (JRW)~\cite{JRW}, and this bound is tight in the sense that 
there exists a pure-state ensemble which saturates the inequality.

\vspace{1mm} {\em Proof}: If the initial ensemble $\varepsilon$ is 
pure, then the final ensembles $\varepsilon_{j}$ are also pure.  As a 
result, the accessible information of each of these ensembles is 
bounded below by $Q[\rho'_{j}]$~\cite{JRW}.  We can therefore replace 
the final sum in Eq.(\ref{accb}) by $\sum_{j}Q_{j}Q[\rho'_{j}]$, which 
gives Eq.(\ref{bsub}).

That the bound can be achieved can be shown by calculating $\Delta 
I_{\mbox{\scriptsize i}}$ for the uniform ensemble over pure states, 
being the unique distribution over pure states which is invariant 
under unitary transformations.  In this case the ensemble state is 
given by
\begin{equation}
 \rho = \int \! |\psi\rangle \langle\psi| \,\, d|\psi\rangle = 
 \frac{I}{N} ,
\end{equation}
where $d|\psi\rangle$ represents integration over the unitarily 
invariant, or {\em Haar}, measure~\cite{Jones}.  The accessible 
information is given by $Q[I/N]$~\cite{JRW}.  The information obtained 
by a general measurement may be calculated directly:
\begin{eqnarray}
  \Delta I_{\mbox{\scriptsize i}} & = &  H[Q_{j}] - \int H[Q(j||\psi\rangle)] 
                                                         d|\psi\rangle 
                                                    \nonumber \\
     & = & \ln N + \sum_{j} \mbox{Tr}[E_{j}] 
           \int \langle\psi|\rho'_{j}|\psi\rangle 
           \ln(\langle\psi|\rho'_{j}|\psi\rangle) d|\psi\rangle 
           \nonumber \\
     & = & Q[I/N] - \sum_{j}Q_{j}Q[\rho'_{j}]                           
\end{eqnarray}
where the integral in the second line is performed using the 
techniques in Ref.~\cite{JRW}.  $\square$

The above result reveals a special property of the uniform ensemble: 
no matter what incomplete measurement is performed on it, the 
information which is not retrieved by the measurement can always be 
extracted by subsequent measurements.  To see this we first use the 
polar decomposition theorem as before to write $A_{j} = U_{j} 
\sqrt{E_{j}}$.  The final state is then 
given by $\rho'_{j} = U_{j}E_{j}U_{j}^{\dagger}/\mbox{Tr}[E_{j}]$.  It 
is convenient to write the final ensemble as a distribution over 
unormalized states, $|\tilde{\phi}_{j}\rangle$.  Writing these states 
in terms of the final state $\rho'_{j}$, we have
\begin{equation}
 |\tilde{\phi}_{j}\rangle = \sqrt{\rho'_{j}} |\psi'_{j}\rangle ,
\end{equation}
where $|\psi'_{j}\rangle = U_{j}|\psi\rangle$.  The probability 
density of these states in the final ensemble (with respect to the 
Haar measure) is $P(|\tilde{\phi}_{j}\rangle) = 
\langle\tilde{\phi}_{j} |\tilde{\phi}_{j} \rangle$.  Since the 
ensemble of states $|\psi\rangle$ is uniform, so is the ensemble of 
states $|\psi'_{j}\rangle$.  The final ensemble, $\varepsilon_{j} = 
\{P(|\tilde{\phi}_{j}\rangle), |\tilde{\phi}_{j}\rangle \}$, is 
referred to as a `distortion' of the uniform ensemble by the state 
$\rho'_{j}$.  Since JRW have shown that such a distortion has an 
accessible information equal to $Q[\rho'_{j}]$, all the information 
missing in the incomplete measurement is accessible in the final 
ensembles $\varepsilon_{j}$.

While the dual Holevo bound is saturated for pure-state ensembles 
which {\em maximize} the accessible information (and measurements 
whose operators commute with the ensemble state), the bound given by 
Eq.(\ref{bsub}) is saturated by pure-state ensembles which {\em 
minimize} the accessible information.

Since we have considered so far only efficient quantum measurements, we complete our discussion by examining classical measurements and inefficient quantum measurements. For this purpose it is best that we first introduce the latter. Inefficient measurements are simply efficient measurements in which the observer knows only that one of a subset of the possible results was obtained. As a result, the observers final state of knowledge is given by averaging over a subset of the states $\rho'_{j}$. Thus, if we now label the measurement results by two indices, $k$ and $l$, then in general we can write the actual final states for an observer who makes an inefficient measurement as $\tilde{\rho}_{k} = \sum_{l} A_{kl}\rho A^{\dagger}_{kl} / Q_{k}$, where $Q_{k}$ is the probability that the final state is $\tilde{\rho}_{k}$.

Now, classical measurements are described by the subset of quantum measurements in which all the encoding states, $\rho_{i}$, and all the measurement operators, $A_{j}$, are mutually commuting (for a discussion, see e.g.~\cite{pool}). As a result it is easily shown that inefficient classical measurements are merely efficient classical measurements, and thus Eq.(\ref{mb}) remains true for all classical measurements. In fact, if the encoding states are {\em pure} classical states (i.e. individual classical states rather than distributions), then the bound is always saturated with equality.

For inefficient quantum measurements however, Eq.(\ref{mb}) does {\em not} hold. The reason for this is that for inefficient measurements $\Delta I_{\mbox{\scriptsize f}}$ can be negative (whereas $\Delta I_{\mbox{\scriptsize i}}$ is always non-negative). An example of such a  situation is one in which the initial state $\rho$ is not maximally mixed, and the observer performs a von Neumann measurement in a basis unbiased with respect to the eigenbasis of $\rho$. If the observer has no knowledge of the outcome, then her final state is maximally mixed. Further, if one mixes, (in the sense of \cite{mixing}) this measurement with one whose measurement operators commute with $\rho$, it is not hard to obtain a measurement in which both $\Delta I_{\mbox{\scriptsize i}}$ and $\Delta I_{\mbox{\scriptsize f}}$ are positive, but which violates Eq.(\ref{mb}).

The author is grateful to Michael J. W. Hall for helpful discussions.

\end{document}